\documentclass[pra,twocolumn]{revtex4}
\newcommand{\be}{\begin{equation}}
\newcommand{\ee}{\end{equation}}
\newcommand{\ben}{\begin{equation*}}
\newcommand{\een}{\end{equation*}}
\newcommand{\bea}{\begin{eqnarray}}
\newcommand{\eea}{\end{eqnarray}}
\newcommand{\bean}{\begin{eqnarray*}}
\newcommand{\eean}{\end{eqnarray*}}

\usepackage{epsfig}

\begin{document}
\title{IR control of atoms laser instability} \author{L.M.Castellano and H.Vinck}
\affiliation{{\small Instituto de F\'{\i}sica, Universidad de Antioquia, A.A. 1226}\\
Medell\'{\i}n-Colombia}

\begin{abstract}
In this paper we show how to control the quantum laser atoms
instability using IR radiation. The control  can be achieved by
controlling the scattering length constant via the infrared
coupling constant. This method is applied  in the scheme of a
continuous CW laser and involves three occupation levels in the
condensed atoms. This atoms in Lambda configuration  the
description of which is given by a Reformed Gross-Pitaevskii
equation (RGPE), together with a rate equation. The system is
taken to a nonconservative complex Ginzburg Landau equation (CGLE)
description from where we use the Benjamin-Feir stability
criterium. This method allows us a theoretical construction of any
atom laser even for negative interaction constant as is the case
of $^{7}Li$.
\end{abstract}
\maketitle

Since the making of the first condensed \cite{buo} and the later
study of its coherence properties \cite{ket1} at least two other
systems have been proposed \cite{kas, hag} for the construction of
an atom laser. An atom laser is defined in analogy to optical
laser and in a similar way it will operate in a dynamical steady
state if some threshold requirement is satisfied \cite{holl}. In
fact, a pulsed atom laser version have been already demonstrated
\cite{ket2, brad1, brad2} been constructed by  Hansch \cite{Hans}
and their group. Despite of this the stability control remains a
difficulty to overcome, even from the onset of BEC formation,
specially for those cases of negative scattering length constants
as $^{7}Li$. The main difficulties are the dissipative mechanism
given by spontaneous emission and subsequent reabsorption of
photons. The first is necessary to obtain the required low
temperatures but the second should be avoided because heating is
introduced in the system. Several possible solutions  both,
dynamical and geometrical have been proposed for the reabsorption
problem. All the geometrical proposals are based on the reduction
of the dimensionality of the trap \cite{ols, pfau} while dynamical
ones rely almost entirely upon optical pumping mechanisms
\cite{sat, san}. In this paper we use a non adiabatic approach in
which we consider two IR fields with frequency $\omega_{1}$ and
$\omega_{2}$ coupling the levels $ 1\Leftrightarrow 3 $ and
$2\Leftrightarrow3$ respectively, as in Fig 1.\\

\begin{figure}[h]
\centering \epsfig{file=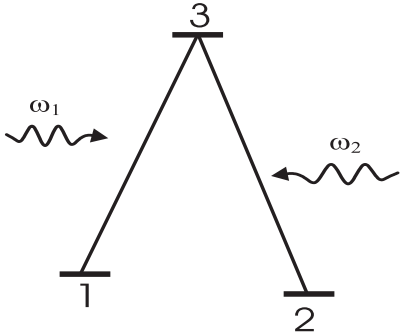,height=125pt, width=150pt}
\caption{The three level atom in interaction with IR Radiation.
$w_{1}$,$w_{2}$, are the coupling and probe field respectively.}
\end{figure}

In this framework atoms in the ground state $|1\rangle$ are in the
optically precooled reservoir before pumped to the MOT trap, where
the atoms will be excited into the  state $|3\rangle$ by using IR
$w_{1}$ field. This corresponds to the thermal cloud or
uncondensed phase. The state $|2\rangle$ corresponds to the
trapped state that will be evaporative cooled e.g  previous to the
condensed phase. The wave intensity and field detunnings should be
properly adjusted in order to the temporal windowing of the
$1\Leftrightarrow 3 $ transition during a time $ t_d =
1/\Delta \nu $, with $\Delta\nu = \nu_{1}- \nu_{2}$ and
$\nu_{1}$, $\nu_{2}$  being the frequencies of fields  1 and 2
respectively. We are therefore quenching one transition
($1\Leftrightarrow 3 $) during a time $t_d$ and allowing the other
one ($2\Leftrightarrow3$) with the appropriate value of
$\Omega_{32}$.\\

In the method proposed the influence of reabsorption mechanism is
then minimized by adjusting the dark state window at will in such
a way that $\Gamma_{32}\ll\Omega_{32}$. $\Gamma_{32}$,
$\Omega_{32}$ are the spontaneous decay rate from level $3$, and
Rabi flopping frequency associated to the probe field
respectively. We point out that this is a different picture than
considered in that of ``Festina Lente" limit \cite{cir} since in
that method the comparison is done between the trap
frequency $w$ and fluorescence rate $\gamma $.\\

Note that, with this mechanism we propose loading the thermal
cloud into the magnetic trap directly and continuously, hence
carrying out the the evaporative cooling in a steady state regime
and thus a CW atom laser is obtained.\\

We are then  left with the atomic pump from the uncondensed
$|3\rangle$ level to the phase condensed state $|2\rangle$
described by the hamiltonian
\be
H_{i}=g_{2}V(b_{2}a_{2}^{\dagger}\phi_{2}^{\dagger}+b_{2}^{\dagger}a_{2}\phi_{2}),
\ee

\noindent where $g_{2}$ is the IR coupling constant in the unit of volumen  defined as $g_{2}=\mu_{23}\sqrt{\frac{\hbar\omega_{2}}{\epsilon_{0}V}}.$ $b_{i}$ ($b_{i}^{\dagger}$) are the annihilation
(creation) operators for $\omega_{i}$ photons and 
$a_{i}$ ($a_{i}^{\dagger}$) are the annihilation (creation)
operators for atoms in the internal state $i$ defined in the usual
matter field operator  as
$\psi(\overrightarrow{r})=\sum_{\alpha}\phi_{\alpha}(\overrightarrow{r})a_{\alpha}$
where $\phi_{\alpha}(\overrightarrow{r})$ are the spatial
dependent particle wave function and $V$ is the volume.
\\
Following the procedure of Ref. \cite{dalf1}, we calculate  the IR
coupling term with \be
[\phi,H_{i}]=\Omega_{2}V\hbar|\phi_{2}|^{2}, \ee

\noindent where
$\Omega\rightarrow\Omega_{2}=g_{2}b_{2}/\hbar$ stands for
the Rabi frequency, and $V$ as the cavity volume. In the other
hand, since $\phi_{2}$ is the condensed phase, we have dropped
subindex $2$ and substitute $\phi_{2}\rightarrow\phi$; from now on
we replace {$\Omega_{32}\rightarrow\Omega_{2}\rightarrow\Omega
$}.\\

The condensed atoms are described by the GPE 

\be
i\hbar\dot{\phi}=\nabla^{2}\phi+V_{t}\phi+g |\phi|^{2}\phi,
\ee

\noindent
where $\phi$ is the condensed wave function, $V_{t}$ is
the harmonic trapping potential and $g$ the interatomic
interaction constant given by $g = 4\pi \hbar^{2}a/m$. The
scattering length constant is $a$ and $m$ the atomic mass . We
adopt the generic model for an atom laser described by Kneer
{\emph et al} in ref \cite{wolf} and rewrite equation (1) with
terms of sink and gain (pumping and outcoupling) plus the IR
coupling giving by expression 2. We then couple the resulting
equation with a rate equation as proposed by Spreeuw {\emph et al}
\cite{speew},
\bea
i\hbar\dot{\phi}&=& \nabla^{2}\phi+V_{t}\phi + g|\phi|^{2}\phi \nonumber\\
&&+ \frac{i}{2}\hbar\Gamma n_{u}\phi -
\frac{i}{2}\hbar\gamma_{c}\phi +\Omega V\hbar|\phi|^{2}\phi.\\
\dot{n_{u}}&=&R(r)-\gamma_{u}n_{u}- \Gamma n_{c}n_{u}, \eea

\noindent Here we prefer to deal with a local coupling. $\Gamma$
is the local $(m^{3}s^{-1})$ rate constant coupling the condensed
field $\phi$ with the uncondensed density $n_{u}$, and is related
to the global rate used in Ref. [16]
$\Gamma^{'}=\frac{\Gamma^{´}}V(s^{-1})$. $\gamma_{c}$ is the
escape rate of condensed atoms leaving the BEC (outcoupled atoms)
and $\gamma_{u}$ the escape rate for the uncondensed atoms (its
inverse plays as the life time of the trap).
$n_{c}=|\phi|^{2}$ is the local density of the condensed
atoms. $R(r)$ is the non uniform, locally depleted pumping process
for the uncondensed atoms coupled to the condensed phase. Because
of this local depletion, the uncondensed phase $n_u$ has a
sensible space dependence and undergoes a diffusion process
\cite{luis}. Beside this, the influence of $R $ upon BEC dynamics
have been shown to be critical at high pumping rates, where
three-body recombination suppress the high frequency dynamics
which is an important factor that limit the atom laser stability \cite{olga}.\\

Now we have to add a real space constant diffusion term
$D_r\nabla^{2}n_u$ to equation $(5)$ and after applying an
adiabatic elimination procedure \cite{Haken}, find a quasi
stationary solution for $n_u$ in terms of $\phi$ and replace it
into equation $(4)$, which then becomes a closed equation for
$\phi$. All this procedure is done on the scenario for which the
$\phi$ dynamics is much slower compared to $n_u$ dynamics.\\

Following the procedure of Ref. \cite{luis}, we have \be
n_{u}=\frac{R}{\gamma_{u}}(1+\frac{D_{r}}{\gamma_{u}}\nabla^{2}-\frac{\Gamma
}{\gamma_{u}}|\phi|^{2}). \ee

\noindent As we replace this expression in Eq.(4), the operator
$\nabla^{2}$ acts on its right upon the space function $\phi$. By
doing this we derive a closed equation for $\phi$:

\bea \dot{\phi}&=& [\frac{\Gamma
RD_{r}}{2\gamma_{u}^{2}}+\frac{i\hbar}{2m}]\nabla^{2}\phi\
-[\frac{\Gamma^{2}R}{2\gamma_{u}^{2}}+i(\Omega V+\frac{g}{\hbar})]|\phi|^{2}|\phi| \nonumber \\
&+& [(\frac{\Gamma
R}{2\gamma_{u}}-\frac{\gamma_{c}}{2})-i\frac{V_{t}}{\hbar}]\phi.
\eea

We now write the parameterized CGL Eq.(8)
\be \dot{\phi}=
\varepsilon\phi+(1+ic_{1})\nabla^{2}\phi-(1+ic_{2})|\phi|^{2}\phi,
\ee

\noindent where
\bea
\varepsilon &=&
\frac{1}{2}(\frac{R\Gamma}{\gamma_{u}\gamma_{c}}-1),\\
c_{1}&=& \frac{\hbar\gamma_{u}^{2}}{mRD_{r}\Gamma},\\
c_{2}&=& \frac{2\gamma_{u}^{2}}{R\Gamma^{2}}(\Omega V +
\frac{g}{\hbar}).
\eea

\noindent Eq. (7), is based on sound physical ground, is far from
being a purely conservative (GP) or purely dissipative (real
Ginzburg Landau) Equation. Instead it display both characters.

 The term $-i\frac{V_{t} }{\hbar}\phi = -ic_{0}\phi$ in Eq. (7) have been
eliminated by the rotation transformation $\phi\longrightarrow\phi\,
e^{-ic_{0}t}$. The model have been made dimensionless using
characteristic length and time scales:
$l_{0}=(\frac{RD_{r}\Gamma}{2\gamma_{u}^{2}\gamma_{c}})^{\frac{1}{2}}\simeq
3.10^{-6}\, m $. Thus, dimensionless space coordinates are
now $x/l_{0}$, $y/l_{0}$, $z/l_{0}$.
Furthermore $\tau = \gamma_{c}t$ with $\gamma_{c}=50 s^{-1}$  , and $|\phi|^{2}\longrightarrow
(\frac{ R\Gamma^{2}}{2\gamma_{u}^{2}\gamma_{c}})|\phi|^{2}$.\\

We are now ready to write down, the Benjamin-Feir
criterium \cite{ben} for stability of the open system \be
c_{1}(-c_{2})<1. \ee

Note that $c_{1}$, $c_{2}$ displays explicit dependence on the
pumping rates and Rabi frequency allowing us a  direct control on
the atom laser stability throughout application of Benjamin-Feir
criterium given by expression (12)
In what follows we consider two cases : \\
a) The $^{87}Rb$ (Fig 2). We have  the experimental reasonable values
$\gamma_{u}=1500 s^{-1}$, $\Gamma=
7.02 \times 10^{-16} m^{3}s^{-1}$, $R=2.13 \times 10^{20} m^{-3}s^{-1}$, $D_{r}
=2 \times 10^{-8} m^{2}s^{-1}$, $g/\hbar=4.8 \times 10^{-17} m^{3}s^{-1}$,$ V=2.5 \times 10^{-16}m^{3}$.
For this case it can be seen that for any IR Rabi coupling the
$^{87}Rb$, Atom laser, is always stable and the only possible
limit is due to the upper values permitted for the $\Omega$.

\begin{figure}[h]
\centering
\epsfig{file=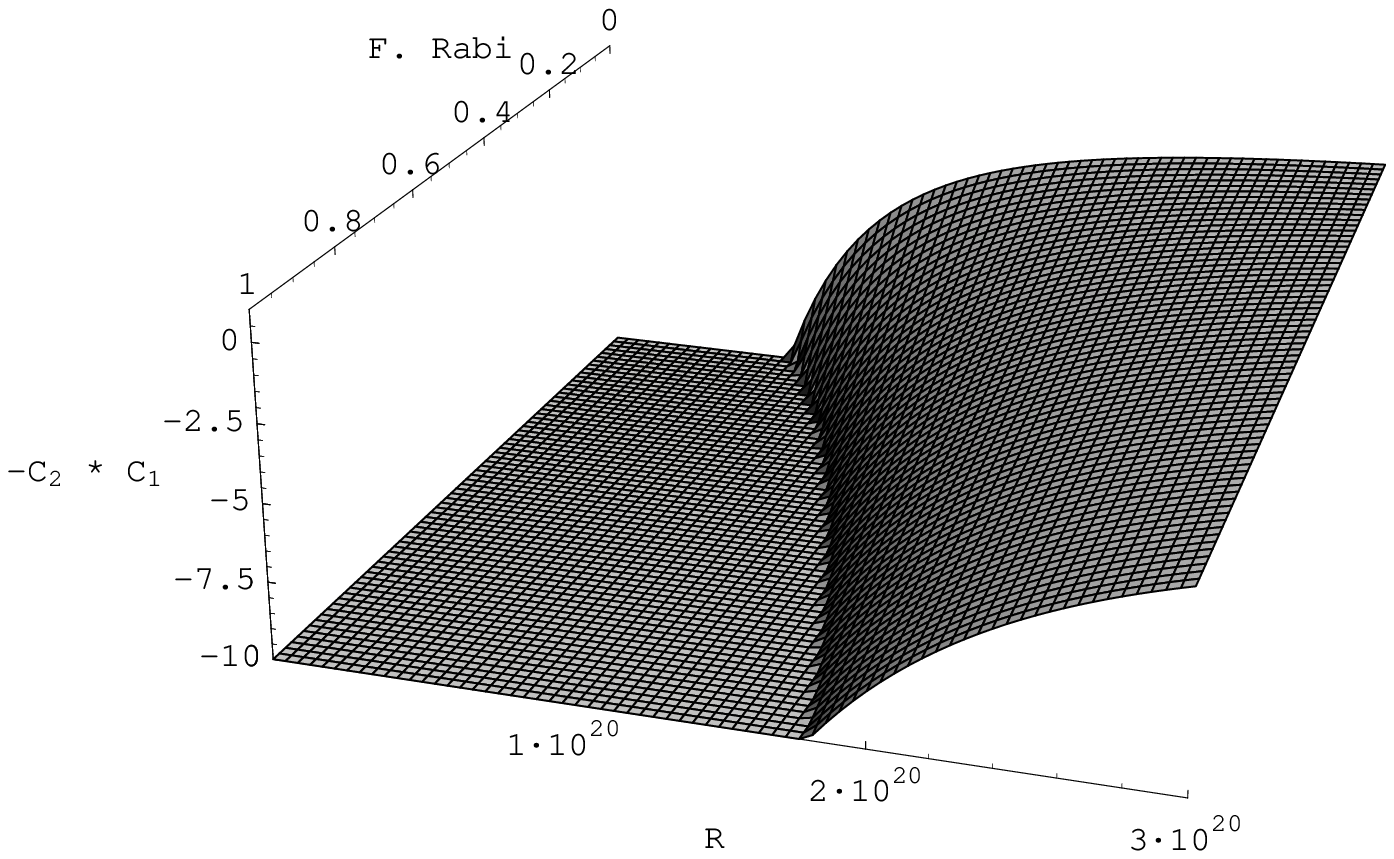,height=150pt, width=200pt}
\caption{The stability condition for $^{87}Rb$ against IR Rabbi frequency and
pumping rate R.}
\end{figure}

b) $^{7}Li$ (Fig 3). We use now the next values: $\gamma_{u}= 87 s^{-1}$, $\Gamma =
5.4 \times 10^{-17} m^{3}s^{-1}$, $R = 1.61 \times 10^{20} m^{-3}s^{-1}$, $D_{r} =
2 \times 10^{-8} m^{2}s^{-1}$, $g/\hbar=
-1.61 \times 10^{-16} m^{3}s^{-1}$, $V=2.5 \times 10^{-16} m^{3}$.\\

\begin{figure}[h]\centering%
\epsfig{file=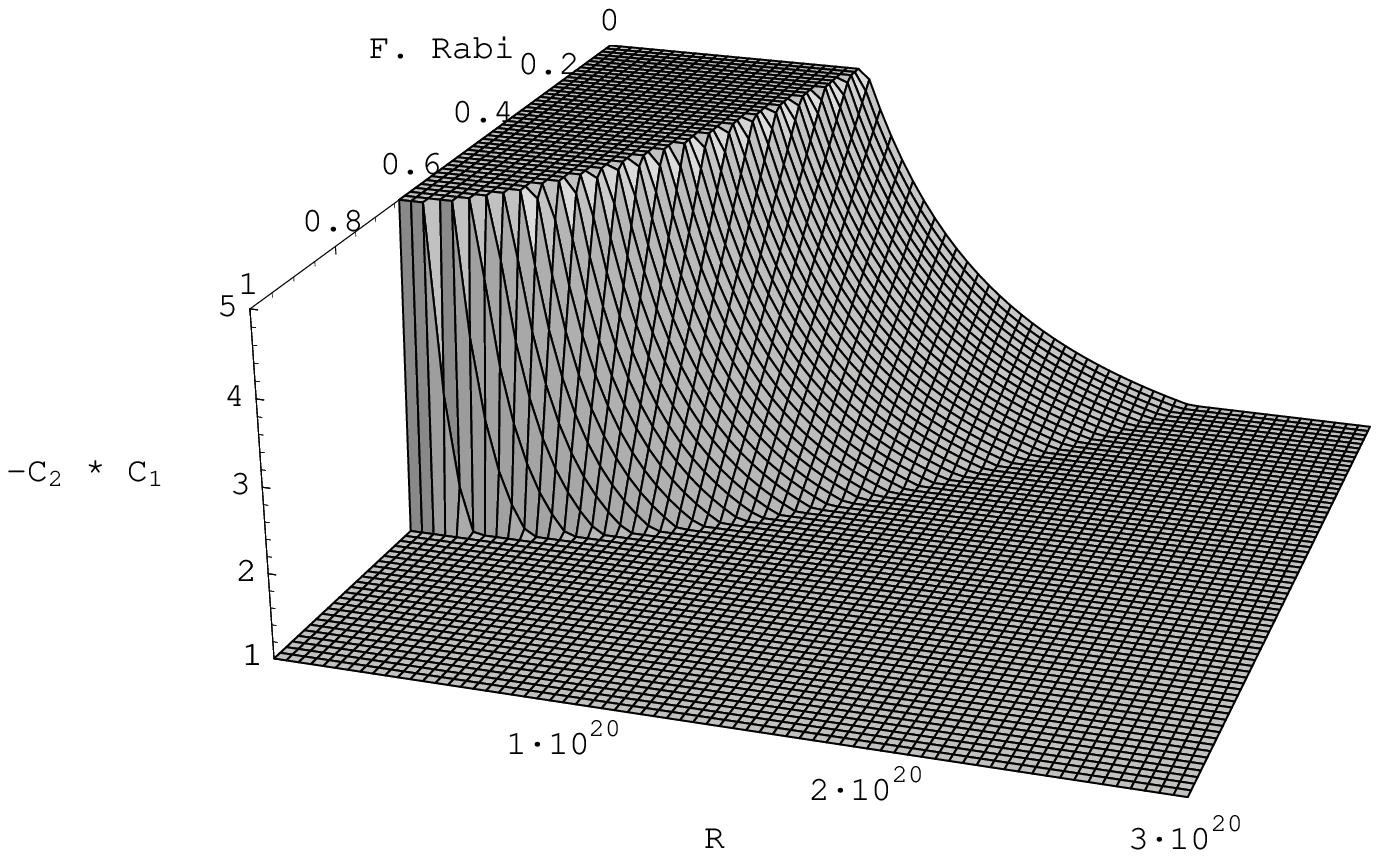,height=150pt, width=200pt} \caption{The
instability condition for litio against IR Rabbi frequency and
pumping rate R}
\end{figure}

\begin{figure}[h]\centering
\epsfig{file=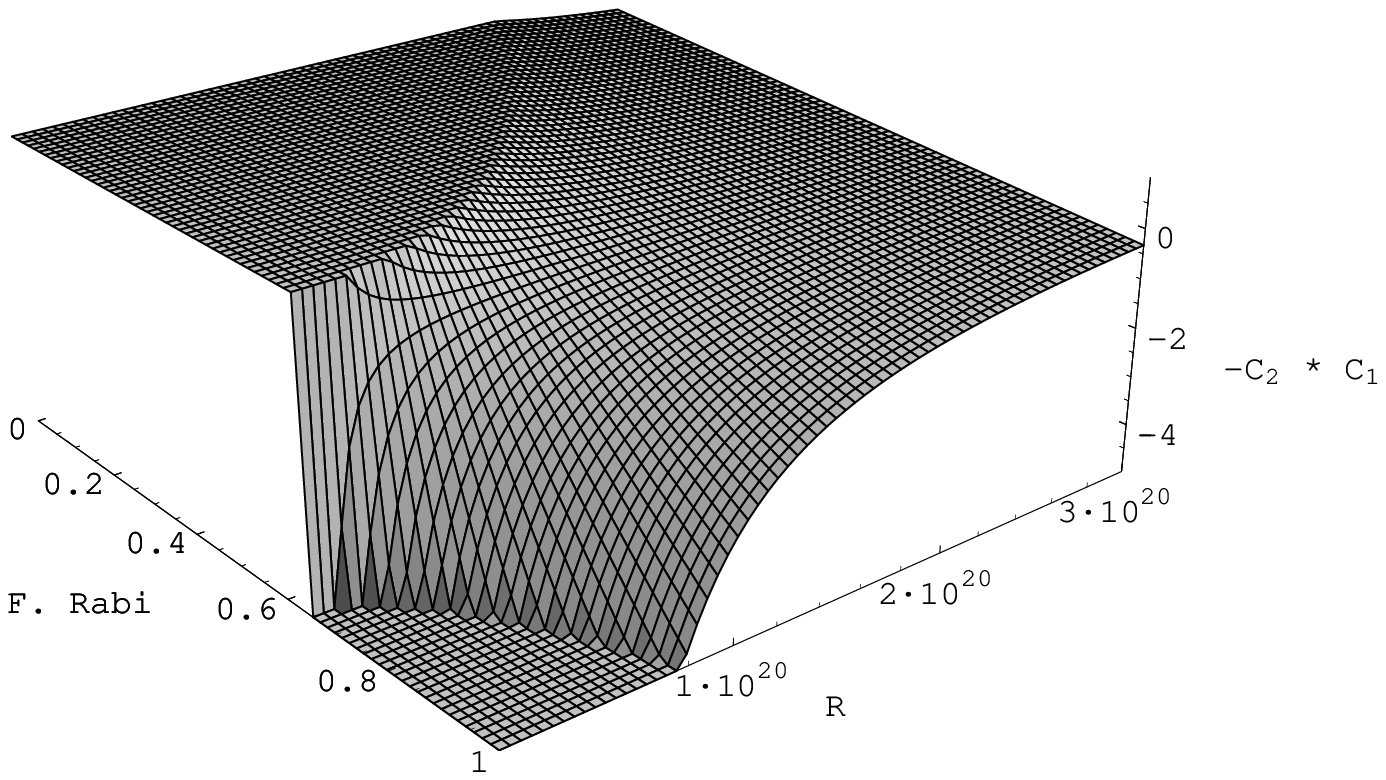,height=150pt, width=200pt} \caption{The
stability condition for litio against IR Rabi frequency and
pumping rate R}
\end{figure}

Note that in this case  there is some range of Rabi frequency for
non stability, running for the values $0<\Omega< 0.35 $. In
the other hand  there is an interesting range of values for
stability as displayed in the Fig. 2 from $0.35<\Omega< 0.6$.
Discussion about the upper limit of stability of the atom laser
remains. It comes to our attention a very recent work by Haine et
al \cite{ha}, where strong dependence on pumping rates are
displayed for stability of $^{85}Rb$ atoms. However in this
numerical work there is no IR interaction. In conclusion we have
shown the possibility of controlling the laser atoms stability in
the framework of quasi- CW optical pumping and IR coupling , even
for those cases of negative scattering length as $^{7}Li$. As a
final comment, we point out the wide spectrum of theoretical
possibilities for studies of atoms laser  using the CGLE  which
seems to be a properly equation, for describing the dynamical of
an open BEC.


\begin{thebibliography}{99}

\bibitem{buo} M. H. Anderson, J. R. Ensher, M. R. Matthews, C. E. Wieman,E. A. Cornell, Science
{\bf{269,}} 198 (1995).
\bibitem{ket1} M. O. Mewes, M. R. Andrews, D. M. Kurn, D. S. Durfee, C. G. Townsend, and
W. Ketterle, Phys.Rev.Lett. {\bf{78,}} 582 (1997).
\bibitem{kas} B. P. Anderson and M. A. Kasevich, Science {\bf{282,}} 1686
(1998).
\bibitem{hag} E. W. Hagley et al, Science {\bf{283,}} 1706 (1999).
\bibitem{holl} J. Williams, R. Walser, C. Wieman, J. Cooper, and
M. Holland, Phys. Rev. A {\bf{57,}} 2030 (1998).
\bibitem{brad1} C. C. Bradley, C. A. Sackett, J. J. Tollet, and R. G. Hulet,
Phys. Rev. Lett. {\bf{75,}} 1687 (1995).
\bibitem{brad2} C. C. Bradley, C. A. Sackett, and R. G. Hulet, Phys. Rev.
Lett. {\bf{78,}} 985 (1997).
\bibitem{ket2} K. B. Davis, M. Mewes, M. R. Andrews, N. J. Van Drutten, D. S. Durfee, D. M. Kurn, and W. Ketterle, Phys.
Rev. Lett. {\bf{75,}} 3969 (1995).
\bibitem{Hans} I. Bloch, T. W. Hansch, and T. Esslinger, Phys. Rev. Lett. {\bf{82,}} 3008
(1999).
\bibitem{ols} M. Olshan, Y.Castin, and J.Dalibard, Proc.$12^{th}$Int. Conf. on Laser Spectroscopy, M. Inguscio,
M. Allegrini, and A. Lasso, Eds.(World Scientific, Singapore,
1996).
\bibitem{pfau} T. Pfau, and J. Mlynek, OSA Trends in Optics and Photonics Series {\bf{7,}} 33
(1997).
\bibitem{sat} S. Bhongale and M.Holland, quant-Ph/0007039v1.
\bibitem{san} L. Santos, F. Floegel, T. Pfau and M. Lewenstein,
quant-ph/0007003v1.
\bibitem{cir} J. I. Cirac, M. Lewenstein, and P. Zoller, Europhys. Lett. {\bf{35,}} 647
(1996).
\bibitem{dalf1} F. Dalfovo, S. Giorgini, L. P. Pitaevskii, S. Stringari, Rev. Mod. Phys. {\bf{71,}} 463
(1999).
\bibitem{wolf} B. Kneer, T. Wong, K. Vogel, W. P. Schleich, D. F. Walls, Phys. Rev. A
{\bf{58,}} 4841 (1998).
\bibitem {speew} R. J. C. Spreeuw, T. Pfau, U. Janicke, M. Wilkens,
Europhys. Lett. {\bf{32,}} 469 (1995).
\bibitem{luis} F. T. Arecchi, J. Bragard, L. M. Castellano, Optics
Communications {\bf{179,}} 149 (2000).
\bibitem{olga} N. Robins, C. Savage, and E. A. Ostrovskaya, Phys. Rev. A
{\bf{64,}} 043605 (2001).
\bibitem{Haken} H. Haken, {\emph Advanced Synergetic} (Springer, Berlin,
1983).
\bibitem{ben} T. B. Benjamin, J. E. Feir, J. Fluid Mech. {\bf{27,}} 417
(1967).
\bibitem{ha} S. A. Haine, J. J. Hope, N. P. Robins, and C. M.
Savage, Phys. Rev. Lett. {\bf{88,}} 170403(2002)
\end{thebibliography}
\end{document}